# A STATISTICAL APPROACH TO SIMULTANEOUS MAPPING AND LOCALIZATION FOR MOBILE ROBOTS[1]

BY ANITA ARANEDA[1], STEPHEN E. FIENBERG[2] AND ALVARO SOTO[1]

*Pontificia Universidad Católica de Chile, Carnegie Mellon University and Pontificia Universidad Católica de Chile*

Mobile robots require basic information to navigate through an environment: they need to know where they are (localization) and they need to know where they are going. For the latter, robots need a map of the environment. Using sensors of a variety of forms, robots gather information as they move through an environment in order to build a map. In this paper we present a novel sampling algorithm to solving the simultaneous mapping and localization (SLAM) problem in indoor environments. We approach the problem from a Bayesian statistics perspective. The data correspond to a set of range finder and odometer measurements, obtained at discrete time instants. We focus on the estimation of the posterior distribution over the space of possible maps given the data. By exploiting different factorizations of this distribution, we derive three sampling algorithms based on importance sampling. We illustrate the results of our approach by testing the algorithms with two real data sets obtained through robot navigation inside office buildings at Carnegie Mellon University and the Pontificia Universidad Catolica de Chile.

**1. Introduction.** Mobile robots require basic information to navigate through an environment: they need to know where they are (localization) and they need to know where they are going. For the latter, robots need a map of the environment. Using sensors of a variety of forms, robots gather information as they move through an environment in order to build a map. There are many algorithmic approaches to deal with this problem; for example, see the discussion in [4]. In this paper we examine data gathered by

Received January 2007; revised April 2007.

[1]Supported by Fondecyt Grant 1050653 at the Pontificia Universidad Católica de Chile.

[2]Supported in part by NSF Grant SES-9720374 to Carnegie Mellon University and NSF Grant DMS-04-39734 to the Institute for Mathematics and Its Application at the University of Minnesota.

Supplementary material available to http://imstat.org/aoas/supplements

two mobile robots, executing a traversal through two different office environments, using an odometer and a simple set of laser readings from sensors. In the past, the processing of such data has benefited enormously from a probabilistic approach that attempts to use the data to form estimates and density functions of the basic quantities of interest [4, 10, 24, 26].

The literature on "probabilistic robotics" has focused heavily on the problems of localization, knowing precisely where the robot is, and of mapping the environment. These are intertwined, that is, to build a map of an environment, the robot needs to know the locations it has visited, but knowing the locations require knowledge of a map. Therefore, the probabilistic robotics problem involves the performance of these dual tasks and is known as Simultaneous Mapping and Localization (SLAM) [16]. It is natural to think of addressing SLAM using a Bayesian approach which puts a posterior distribution over the space of all possible maps and then updates the distribution using the information that the robot acquires as it moves through the environment. This Bayesian solution in some sense maximizes the information available for SLAM [5]. Most of the robotics literature on SLAM utilizes a variety of approximations that allow for real-time calculations and updating and thus, of necessity, simplifies this Bayesian conceptual formulation of the SLAM problem.

Our first data set comes from an experiment conducted with a mobile robot, Pearl, at Carnegie Mellon University in Wean Hall (see Figure 1). Our second data set comes from a second robot, this one navigating inside the Computer Science Department at the Pontificia Universidad Catolica de Chile. Both data sets consist of a set of noisy measurements obtained by an odometer and a laser range finder mounted on the robot. Odometer readings convey information about the robot's relative location. They correspond to rotational and translational measures of the robot movements. Laser readings convey information about the location of landmarks, with respect to the robot's location. They correspond to a set of scalar quantities indicating the distances from the robot to the nearest obstacle in a set of previously specified directions.

Using this type of data, we propose a complete probabilistic representation of the SLAM problem and obtain a Bayesian solution. We formalize the problem of mapping as the problem of learning the posterior distribution of the map given the data. Our key idea is based on noting that the posterior distribution of the map is determined by the posterior joint distribution of the locations visited by the robot and the distances to the obstacles from those locations. We derive expressions for this posterior joint distribution of locations and distances and show that there is no closed form for it. By exploiting different factorizations of this distribution, we derive three sampling algorithms based on importance sampling.

The outline of the paper is as follows. In Section 2 we discuss previous research in SLAM and compare this literature with our approach. In Section 3 we describe in detail the dependencies and models that define our probabilistic approach. In Section 4 we explore the posterior distribution of maps and develop three sampling algorithms. In Section 5 we apply these algorithms to the data sets collected in Wean Hall Building at Carnegie Mellon University, and in the Computer Science Department at the Pontificia Universidad Catolica de Chile. Finally, in Section 6 we discuss briefly extensions of our model and methodology, in order to allow for both real time implementation and more elaborate forms of data input.

**2. Previous SLAM approaches.** Although there is an extensive robotics research literature dealing with mapping or localization for mobile robots (e.g., see [2, 8, 24, 26]) the SLAM problem is a relatively newer research area, where most efforts have been made over the last couple of decades. An important family of approaches to SLAM is based on versions of the Kalman filter. The pioneering development in this area was the paper by Smith et al. [22] which proposed a basic version of the Hidden Markov Model (HMM) approach widely used today, and then used the Kalman filter to address the problem of estimating topological maps. They assumed a fixed number of landmarks in the environment where these landmarks can be identified by their cartesian coordinates. At a fixed point in time, the set of landmarks coordinates and the location of the robot are assumed to be unobservable or latent variables. As in the Kalman filter, the main assumption is that the posterior distributions of all these variables are Gaussian and that the observations, given the latent variables, can be described by a linear function and a white noise term.

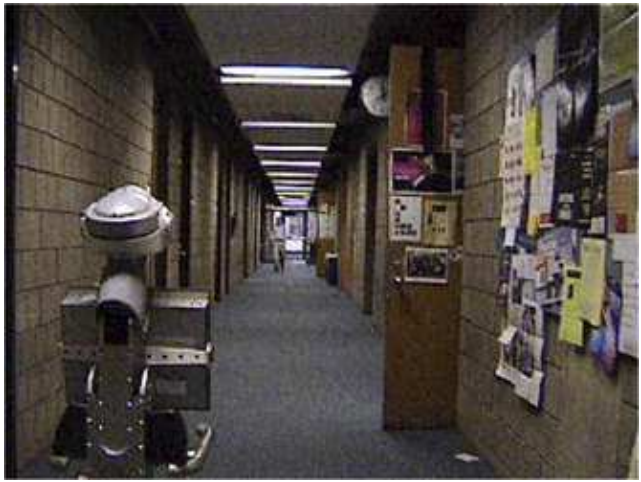

Fig. 1. *The mobile robot Pearl gazing down the corridor of Wean Hall, Carnegie Mellon University. Courtesy of Sebastian Thrun from a video available at* http://robots.stanford.edu/videos.html.

These two assumptions, Gaussian variables and linearity, are somewhat restrictive. The Gaussian assumption makes this approach unsuitable for multimodal distributions that arise when the location of the robot is ambiguous. The linearity assumption is not met in general, since the relation between odometry and locations involves trigonometric functions. The Extended Kalman Filter (EKF) [7, 15] partially handles nonlinearity using a Taylor series approximation.

For the non-Gaussian case, Thrun et al. [25] outlined a general approach that can be used with general distribution functions. Under this approach, however, maximum likelihood estimation is too expensive computationally. As an alternative, Thrun [24] presented an application of the Expectation–Maximization (EM) algorithm [6] applied to mapping. He treated the map as the parameter to be estimated and the locations as part of an HMM, maximizing the expected log likelihood of the observations and the locations, given the map.

A more recent and successful approach to the SLAM problem is the Fast-SLAM algorithm [17]. This approach applies to topological maps, and is based on a factorization of the posterior distribution of maps and locations. For full details on both the models and their algorithmic implementation, see [18]. The key factorization of maps and locations is not part of the model we present here and should be thought of as an approximation which allows for real-time implementation.

Hähnel et al. [13] present an approach that is also based on the description in [24], but this one applied to occupancy grids. This approach finds locations iteratively over time. At each point in time, the algorithm estimates the location visited by the robot as the location that maximizes the probability of the current data, given past data and previous location estimates. The next step finds the map, as the map that maximizes the posterior probability of the estimated locations and the observed data.

Our mapping approach applies to occupancy grid maps of static environments. Our formulation of the problem is based on the approach initially described in Thrun et al. [25]. We build a graphical representation of that formulation where the locations are considered unobservable variables determining the observed odometer readings and, together with the map, determining the observed laser readings. The probability model for the entire process is determined by motion and perception models and a prior distribution for the map.

In contrast with most of these other approaches, we provide a formal probabilistic description of the entire process and develop a Bayesian solution with the goal of estimating the posterior distribution of the map using simulation. The fact that our approach uses a more general motion model than the one used by the Kalman filter and EKF approaches makes it applicable to a wider set of problems. The advantage of this method is that it does not provide a single estimate of the map, as the EM-based solution, but it produces multiple maps showing the notion of variability from the expected posterior map. For localization, we obtain a simulation of the locations visited by the robot from their posterior distribution, as an intermediate step while simulating maps.

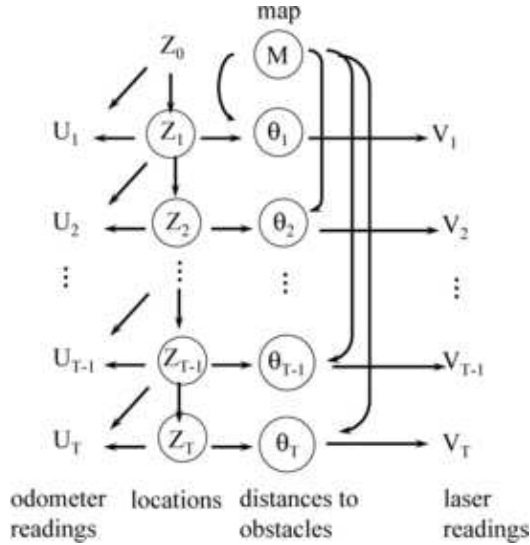

Fig. 2. *Graphical representation of the robotics SLAM problem. In our application at Carnegie Mellon University the robot obtains a total of $T = 3354$ readings.*

**3. Formalization of the problem.** Figure 2 provides a graphical representation of the SLAM problem, where nonobservable variables have been circled for clarity. This representation forms the basis of our model and has some similarities to a suggested representation in an earlier paper by Murphy [19], as well as to probabilistic graphs of the sort we could formulate related to Kalman filter approaches. Our representation was developed in earlier unpublished work and has been implicitly adopted in [26].

In Figure 2 time increases as we go down in the figure. The robot starts at location $Z_0$, and moves to its final location $Z_T$. In the figure $U_t$ correspond to odometer data recorded at time $t$, expressed as differences between locations at times $t$ and $t-1$. The vector $Z_t$ corresponds to the true location visited by the robot at time $t$. Two consecutive locations, $Z_t$ and $Z_{t-1}$, induce an odometer translation reading, $U_t$.

The random variable $\boldsymbol{M} = \{M_{ij}, (i,j) \in \mathcal{I}\}$ represents the map of the environment, where $M_{ij}$ takes the value "$\underline{1}$," if the location $(i,j)$ is occupied by an obstacle, and "0" otherwise, and $\mathcal{I}$ is a suitable set of locations. At each point in time, the map $\boldsymbol{M}$ and a given location of the robot, $Z_t$, determine the distances to obstacles, $\theta_t$. Finally, these distances determine the distribution of the laser readings $V_t$.

According to Figure 2, the distribution of the process is determined by three models. A *motion model* [23] describes the dependency of the current location, $Z_t$, on the previous one and the current odometer reading, $Z_{t-1}$ and $U_t$, respectively. We adopt a Gaussian motion model. A *perception model* [23] describes the dependency of laser readings $V_t$ on the true distances to obstacles, $\theta_t$. We use a truncated Gaussian distribution, with standard deviation $\sigma$ where the limits of the distribution correspond to 0 and $d_{\max}$, the maximum range of the laser device. Finally, for our prior distribution for the map, $\boldsymbol{M}$, we assume that cells in the map are independent, each having the same probability, $p$, of being occupied. Araneda [1] discusses these models in greater detail.

In what follows we use $\boldsymbol{U}$, $\boldsymbol{V}$, $\boldsymbol{Z}$ and $\boldsymbol{\theta}$ to denote the sets of odometer reading differences, laser readings, locations and distances to obstacles, from time 1 to $T$, respectively.

**4. Importance sampling for map inference.** In this section we explore the posterior distribution of maps given the data and derive sampling strategies based on Importance Sampling (IS) [11]. We note that the posterior distribution over the space of possible maps, $\boldsymbol{M}$, is completely determined by the joint posterior distribution of locations and distances to obstacles, $\boldsymbol{Z}$ and $\boldsymbol{\theta}$. Thus, our approaches are intended to explore this last posterior distribution, and they mainly differ in the particular factorization used, this last one being either

$$(4.1) \qquad P(\boldsymbol{Z}, \boldsymbol{\theta} | \boldsymbol{U}, \boldsymbol{V}) = P(\boldsymbol{Z} | \boldsymbol{U}, \boldsymbol{V}) P(\boldsymbol{\theta} | \boldsymbol{Z}, \boldsymbol{U}, \boldsymbol{V}),$$

or

$$(4.2) \qquad P(\boldsymbol{Z}, \boldsymbol{\theta} | \boldsymbol{U}, \boldsymbol{V}) = P(\boldsymbol{\theta} | \boldsymbol{U}, \boldsymbol{V}) P(\boldsymbol{Z} | \boldsymbol{\theta}, \boldsymbol{U}, \boldsymbol{V}).$$

Our first algorithm, based on factorization (4.1), approximates the posterior distribution of the locations by discarding laser readings. From the application of this algorithm to the Wean Hall data set, we learn that odometry alone does not help the robot to recover from odometry error.

Our second algorithm, based on factorization (4.2), approximates the posterior distribution of distances to obstacles by discarding odometer readings. The application of this algorithm brings upfront the problem posed by the many restrictions of the sampling space. We are not able to handle these restrictions analytically and, thus, we sample over broader sampling spaces.

Our main algorithm, also based on factorization (4.2), corresponds to a partially probabilistic algorithm, where the restrictions of the sampling space are relaxed, allowing observations to lie outside the restrictive domain of the distributions. This algorithm is successful in recovering from odometry error by using all odometer and laser information when sampling locations.

In the next sections we describe the derivation of the algorithms. The first algorithm is described briefly, and the details of this derivation are shown in the Appendix. As the second algorithm corresponds to the basis of our main algorithm, we show its derivation in more detail. We finally refer to our main algorithm. The results we mention above are based on the application of the algorithms that we describe in Section 5.

4.1. *Derivation of the first algorithm.* The main feature of the first algorithm is the approximation of the posterior distribution of the locations by the distribution implied by the motion model. In other words, this approach considers odometry information only, when sampling locations, and discards the information about the locations that is contained in laser readings. In particular, consider the factorization in (4.1). We approximate the first term in this product, $P(\boldsymbol{Z}|\boldsymbol{U}, \boldsymbol{V})$, by the motion model, $P(\boldsymbol{Z}|\boldsymbol{U})$.

For the second term in the factorization, $P(\boldsymbol{\theta}|\boldsymbol{Z}, \boldsymbol{U}, \boldsymbol{V})$, we find that we need to focus on terms of the form $P(\theta_{tk}|\boldsymbol{Z}, \boldsymbol{V}, \boldsymbol{\theta}^{t-1})$, where $\theta_{tk}$ corresponds to the distance to the obstacle located in the direction of the $k$th beam of the laser device at time $t$, and $\boldsymbol{\theta}^{t-1}$ denotes the matrix containing all distances up to time $t-1$. Working this expression, and under a Gaussian perception model, we obtain

$$
\begin{aligned}
P(\theta_{tk}&|\boldsymbol{Z}, \boldsymbol{V}, \boldsymbol{\theta}^{t-1}) \\
&\propto \frac{\phi((V_{tk} - \theta_{tk})/\sigma)}{\Phi((d_{\max} - \theta_{tk})/\sigma) - \Phi(-\theta_{tk}/\sigma)} P(\theta_{tk}|\boldsymbol{Z}, \boldsymbol{\theta}^{t-1}, \boldsymbol{V}_{(tk)}) \\
&\quad \times I(0 < V_{tk} < d_{\max}) I(\theta_{tk} > 0),
\end{aligned}
\tag{4.3}
$$

where the term $\boldsymbol{V}_{(tk)}$ corresponds to the set of all laser readings except for reading $V_{tk}$. The functions $\phi$ and $\Phi$ correspond to the density and cumulative function of the standard Gaussian distribution, respectively (see the Appendix for details).

Equation (4.3) implies that we can implement an IS algorithm by sampling $\theta_{tk}$ from a Gaussian distribution with importance weights given by

$$
\begin{aligned}
\omega(\theta_{tk}) &= \frac{1}{\Phi((d_{\max} - \theta_{tk})/\sigma) - \Phi(-\theta_{tk}/\sigma)} P(\theta_{tk}|\boldsymbol{Z}, \boldsymbol{\theta}^{t-1}, \boldsymbol{V}_{(tk)}) \\
&\quad \times I(0 < V_{tk} < d_{\max}) I(\theta_{tk} > 0).
\end{aligned}
\tag{4.4}
$$

The distribution $P(\theta_{tk}|\boldsymbol{Z}, \boldsymbol{\theta}^{t-1}, \boldsymbol{V}_{(tk)})$ in (4.4) can be approximated by a truncated geometric distribution, that we denote by $Tr.Geom(C_{tk}, p)$. See the Appendix for an outline of the derivation. The important consideration in obtaining this result corresponds to the fact that the sampling space of $\theta_{tk}$, given $\boldsymbol{Z}^{t-1}$, $\boldsymbol{\theta}^{t-1}$ and $Z_t$, $C_{tk}$, gets narrower as time increases. We say that the values in this space are *consistent* with the values of $\boldsymbol{Z}^{t-1}$, $\boldsymbol{\theta}^{t-1}$ and $Z_t$. The parameter $p$ in the truncated geometric distribution corresponds to the prior probability of a cell of being occupied.

With this approximation in hand, we can go back to equation (4.4), and rewrite the weight of each sampled value $\theta_{tk}$ as

$$
\omega(\theta_{tk}) = \frac{Tr.Geom(C_{tk}, p; \theta_{tk})}{\Phi((d_{\max} - \theta_{tk})/\sigma) - \Phi(-\theta_{tk}/\sigma)}
$$
$$
\times I(0 < V_{tk} < d_{\max}) I(\theta_{tk} > 0).
$$

(4.5)

4.2. *Derivation of the second algorithm.* This algorithm samples distances to the closest obstacles first, and locations afterward. It disregards information carried by odometer readings when explaining distances to obstacles, approximating their posterior distribution by the perception model.

The approximation used in this second algorithm is more accurate than the one used in the first algorithm, since odometry error accumulates over time, while laser error does not. In addition, the high precision of laser sensors produces accurate readings and, thus, highly valuable information. Thus, dropping odometer readings has a smaller impact than dropping laser readings.

In particular, consider the factorization of the posterior distribution given in (4.2). To sample $\boldsymbol{\theta}$ from the first term, $P(\boldsymbol{\theta}|\boldsymbol{U}, \boldsymbol{V})$, we use that

$$
P(\boldsymbol{\theta}|\boldsymbol{U}, \boldsymbol{V}) \approx P(\boldsymbol{\theta}|\boldsymbol{V}) = \prod_{t=1}^{T} \prod_{k=1}^{N} P(\theta_{tk}|V_{tk})
$$

(4.6)
$$
\propto \prod_{t=1}^{T} \prod_{k=1}^{N} P(V_{tk}|\theta_{tk}) P(\theta_{tk})
$$
$$
= \prod_{t=1}^{T} \prod_{k=1}^{N} \frac{\phi((V_{tk} - \theta_{tk})/\sigma)}{\Phi((d_{\max} - \theta_{tk})/\sigma) - \Phi(-\theta_{tk}/\sigma)} (1-p)^{\theta_{tk}-1} p.
$$

Thus, we can implement an IS algorithm by sampling values of $\theta_{tk}$ from a Gaussian distribution, and associating a weight

(4.7)
$$
\omega(\theta_{tk}) = \frac{1}{\Phi((d_{\max} - \theta_{tk})/\sigma) - \Phi(-\theta_{tk}/\sigma)} (1-p)^{\theta_{tk}-1} p,
$$

to each observation.

For the second term in equation (4.2), $P(\boldsymbol{Z}|\boldsymbol{\theta}, \boldsymbol{U}, \boldsymbol{V})$, we have that

$$
P(\boldsymbol{Z}|\boldsymbol{\theta}, \boldsymbol{U}, \boldsymbol{V}) = \prod_{t=1}^{T} P(Z_t|\boldsymbol{\theta}, \boldsymbol{U}, \boldsymbol{Z}^{t-1})
$$

(4.8)
$$= \prod_{t=1}^{T} P(Z_t | \boldsymbol{\theta}, \boldsymbol{U}_t^T, \boldsymbol{Z}^{t-1}).$$

From this expression, applying Bayes' theorem, we get

$$P(\boldsymbol{Z} | \boldsymbol{\theta}, \boldsymbol{U}, \mathbf{V}) = \prod_{t=1}^{T} P(Z_t | \boldsymbol{U}_t^T, \boldsymbol{Z}^{t-1}) \frac{P(\boldsymbol{\theta} | Z_t, \boldsymbol{U}_t^T, \boldsymbol{Z}^{t-1})}{P(\boldsymbol{\theta} | \boldsymbol{U}_t^T, \boldsymbol{Z}^{t-1})}$$

$$= \prod_{t=1}^{T} P(Z_t | U_t, Z_{t-1}) \frac{P(\boldsymbol{\theta} | Z_t, \boldsymbol{U}_t^T, \boldsymbol{Z}^{t-1})}{P(\boldsymbol{\theta} | \boldsymbol{U}_t^T, \boldsymbol{Z}^{t-1})}.$$

Thus, we can implement an IS strategy by sampling locations, $Z_t$, from the motion model, $P(Z_t | U_t, Z_{t-1})$, and assigning weights

$$\omega(Z_t) = P(\boldsymbol{\theta} | Z_t, \boldsymbol{U}_t^T, \boldsymbol{Z}^{t-1}),$$

as the term $P(\boldsymbol{\theta} | \boldsymbol{U}_t^T, \boldsymbol{Z}^{t-1})$ in the denominator does not depend on $Z_t$.

Decomposing $\boldsymbol{\theta}$ over time, as in

(4.9)
$$\omega(Z_t) = \prod_{i=1}^{t-1} P(\boldsymbol{\theta}_i | Z_t, \boldsymbol{U}_{t+1}^T, \boldsymbol{Z}^{t-1}, \boldsymbol{\theta}^{i-1}) \times P(\boldsymbol{\theta}_t | Z_t, \boldsymbol{U}_{t+1}^T, \boldsymbol{Z}^{t-1}, \boldsymbol{\theta}^{t-1})$$

$$\times \prod_{i=t+1}^{T} P(\boldsymbol{\theta}_i | Z_t, \boldsymbol{U}_{t+1}^T, \boldsymbol{Z}^{t-1}, \boldsymbol{\theta}^{i-1}),$$

we identify three types of terms.

When $i$ is smaller than $t$, as in the terms in the first product in (4.9), there is no distance information available at time $t$ to match with $Z_t$, so we can drop the conditioning on $Z_t$. Thus, the terms in this product do not depend on $Z_t$ and we can drop them from the weight. When $i$ is greater that $t$, the terms in the second product become

$$P(\boldsymbol{\theta}_i | U_i, \boldsymbol{M}(\boldsymbol{Z}^{t-1}, \boldsymbol{\theta}^{t-1})),$$

where $\boldsymbol{M}(\boldsymbol{Z}^{t-1}, \boldsymbol{\theta}^{t-1})$ corresponds to the partial map built with information up to time $t-1$. As the location information at time $i$, $U_i$, is random, computing this term requires an additional integration. To avoid this, we approximate the weights in (4.9) by

(4.10)
$$\omega(Z_t) \approx P(\boldsymbol{\theta}_t | Z_t, \boldsymbol{U}_{t+1}^T, \boldsymbol{Z}^{t-1}, \boldsymbol{\theta}^{t-1})$$

$$= P(\boldsymbol{\theta}_t | Z_t, \boldsymbol{M}(\boldsymbol{Z}^{t-1}, \boldsymbol{\theta}^{t-1}))$$

$$= \prod_{k=1}^{N} P(\boldsymbol{\theta}_{tk} | Z_t, \boldsymbol{M}(\boldsymbol{Z}^{t-1}, \boldsymbol{\theta}^{t-1})).$$

The terms inside the product in equation (4.10) correspond to a truncated geometric distribution (see the Appendix for details).

We note that the weight, $\omega(Z_t) = P(\boldsymbol{\theta}_t | Z_t, \boldsymbol{M}(\boldsymbol{Z}^{t-1}, \boldsymbol{\theta}^{t-1}))$, represents the degree of agreement between true distances to obstacles at time $t$, $\boldsymbol{\theta}_t$, and the sampled location, $Z_t$, within the partial map implied by $\boldsymbol{Z}^{t-1}$ and $\boldsymbol{\theta}^{t-1}$.

4.3. *Derivation of our main algorithm.* Our third algorithm builds upon the derivation of the second algorithm. The key difference lies on the weights used when sampling locations $Z_t$. As we show in the application of the second algorithm in Section 5, problems arise with the weights of the locations in (4.10), due to the imposition of consistency between observations. Our third algorithm relaxes the concept of consistency by redefining weights. This induces a new probabilistic model for the process and we have not explored the probabilistic consequences of using the new weights.

Consider a draw at a single time instant $t$, $(Z_t, \theta_t)$. At this point, there are available draws of locations and distances up to time $t - 1$, $\boldsymbol{Z}^{t-1}$ and $\boldsymbol{\theta}^{t-1}$, respectively. Since we are now dealing with a different concept of consistency, the draws obtained up to time $t - 1$ do not necessarily imply a consistent map, in the sense consistency was understood before. That is, there may be conflicting information in previous draws so that a certain cell is determined as empty by some observations and as occupied by others. Thus, we redefine the map up to time $t - 1$ as a probabilistic map that we label $\tilde{\boldsymbol{M}}^{t-1}$.

To understand the definition of $\tilde{\boldsymbol{M}}^{t-1}$, consider each pair $(Z_l, \theta_l)$, $l = 1, \ldots, t - 1$. Every time one of these pairs is drawn, the status of each cell in the map is determined as empty, occupied or unknown. We define the value of cell $(i, j)$ in $\tilde{\boldsymbol{M}}^{t-1}$ as the proportion of times the cell was determined as occupied, over the number of times the cell was determined either occupied or empty. We determine, afterward, that a cell in $\tilde{\boldsymbol{M}}^{t-1}$ is empty if its value is smaller that $\pi$ and is occupied if its value is greater than or equal to $1 - \pi$. From calibration of the algorithm on the Wean Hall data set described in the next section, we use $\pi = 0.2$.

On the other hand, consider that for laser beam $k$, a sampled value $\theta_{tk}$ and a sampled last location $Z_t$ obtained from the motion model, determine that the cell at the distance $\theta_{tk}$ from $Z_t$ in the $k$th direction is occupied, while the cells between that cell and $Z_t$ are empty.

The algorithm considers the degree of agreement between these two sources of information: the partial map $\tilde{\boldsymbol{M}}^{t-1}$ and the information obtained from $\theta_{tk}$ and the sampled value $Z_t$. Two cells, one in $\tilde{\boldsymbol{M}}^{t-1}$ and the same one in the map determined by $\theta_{tk}$ and $Z_t$, agree if they have been assigned the same value by the two sources. Otherwise, they disagree. Unlabeled cells are discarded. We define the weight of $Z_t$ with respect to the $k$th direction,

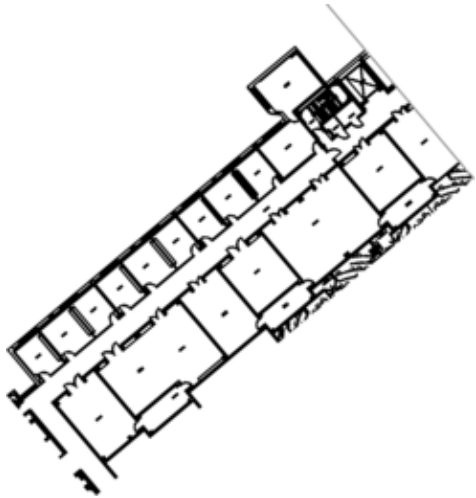

$\omega_k(Z_t)$, as the proportion of cells that agree, over the total number of cells that either agree or disagree. Finally, we define the weight of $Z_t$ as

$$(4.11) \qquad \omega(Z_t) = \prod_{k=1}^{N} \omega_k(Z_t).$$

**5. Implementation and empirical results.**  We first apply the three algorithms described in Section 4 to data obtained from an experimental run in Wean Hall, at Carnegie Mellon University, by Pearl (see Figure 1), a robot equipped with an odometer and a laser sensor. (The data were provided by Nicholas Roy.) Pearl navigated inside the 5th floor of Wean Hall building going back and forth along two corridors shown in the map in Figure 3. Pearl collected data at about 10 recordings per second. In her journey, she took 3354 measurements, each of them consisting in a pair $(U_t, V_t)$, $t = 1, \ldots, 3354$. Her laser sensor sent beams every degree and thus there are 180 distances recorded for each laser reading. The laser sensor had a maximum distance range of $d_{\max} = 10m$.

A map drawn from raw odometer and laser readings is shown in Figure 4. This figure shows how odometry error accumulates so that it seems that Pearl visited three different corridors, instead of two. The smoothness of the depicted walls, however, suggests that error in laser sensor readings is small compared to error in odometer readings.

5.1. *Results of the first algorithm.*  Figure 5 shows a typical path obtained using the first algorithm. We see that large observed rotations in odometry cause large rotations to occur in the sampled path, which explains its curved undesirable shape. Figure 5 shows that paths are unable to recover from the error accumulated by odometry. The lack of ability of the motion model to recover from odometry error shows the need of incorporating the information carried by laser sensor readings when sampling paths, as our second and main algorithms do.

The second step of our sampling process simulates distances to the obstacles from their posterior distribution given the data and the sampled paths. Figure 6 shows the map obtained for the same path shown in Figure 5. The figure shows that laser readings appear to induce larger errors in the sampled maps. This impression is not true, however, if we notice that the sampled locations carry large errors in orientations, which could not be seen

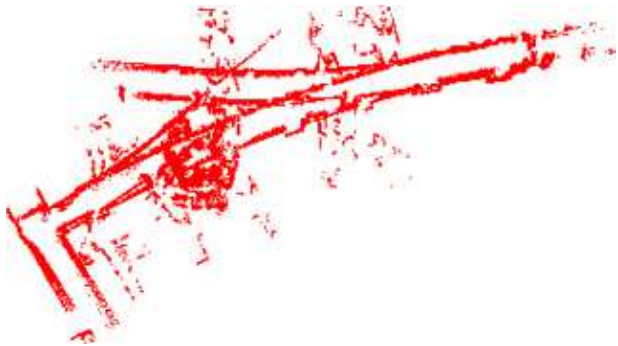

Fig. 4.   *Map obtained from raw data.*

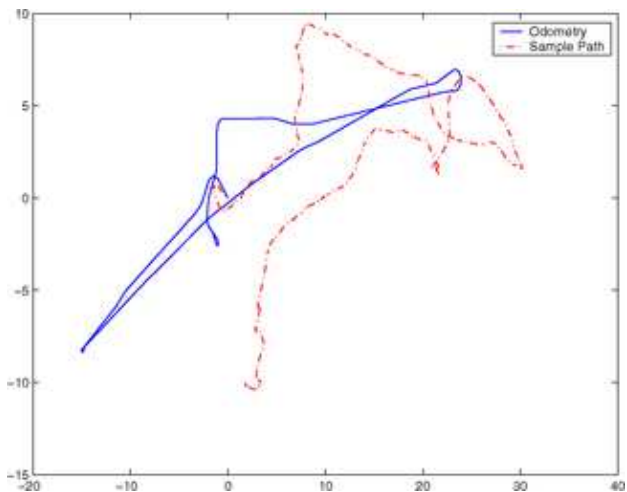

Fig. 5.   *Path sampled by the first algorithm.*

in Figure 5. This causes the obstacles in the map to be located in the wrong places.

5.2. *Results of the second algorithm.* Under this algorithm, we obtain a sample of distances to obstacles $\boldsymbol{\theta}$ from a Gaussian perception model, and assign weights $\omega(\theta_{tk})$ in equation (4.7). It is not possible to visualize the observations obtained in this step within the map, as there are no locations available.

Problems arise early during the second sampling step. When sampling $Z_t$, at instant $t$, there is available a partial map built with information up to time $t-1$, $\boldsymbol{M}^{t-1}(\boldsymbol{Z}^{t-1}, \boldsymbol{\theta}^{t-1})$, and a set of distances, $\theta_t$. At this step, we sample a value of the current location $Z_t$ from the motion model, and assign a weight to this observation that corresponds to the degree of agreement of this location with the previous information.

The partial map that is available, $\boldsymbol{M}^{t-1}$, contains undetermined cells. Thus, for a given location $Z_t$, there is a set of possible distances to the closest obstacles. The weight of the location sampled, $Z_t$, is zero if $\theta_t$ is not in that set. Once $t$ increases and more information is available in $\boldsymbol{M}^{t-1}$, the set of possible distances to obstacles from $Z_t$ gets smaller and, thus, it becomes difficult to draw a value of $Z_t$ with a positive weight. In other words, it becomes hard to find a consistent location $Z_t$.

This issue appears rapidly in the implementation of this algorithm. Once a few locations are sampled, it becomes hard to find consistent subsequent locations, making this implementation unfeasible from the computational point of view.

5.3. *Results of our main algorithm.* We tested our main algorithm with both data sets, from Carnegie Mellon University and from the Pontificia Universidad Catolica de Chile. To get each observation $(\boldsymbol{Z}, \boldsymbol{\theta})$ in the sample, we start by sampling $n$ sets of distances to obstacles, $\theta_{tk}, t = 1, \ldots, T, k = 1, \ldots, N$, from Gaussian distributions, with means $V_{tk}$ and $\hat{\sigma} = 2cm$. Using IS weights according to equation (4.7), we resample one of these observations.

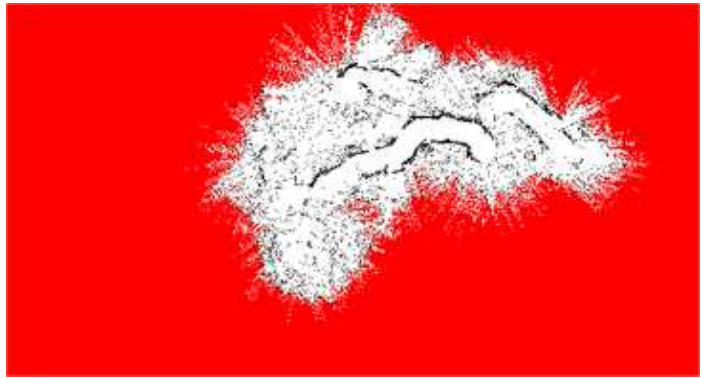

Fig. 6. *Map matching sample path in Figure 5.*

We sample $n$ sets of locations $Z_t, t = 1, \ldots, t$, from the motion model and use the value of $\boldsymbol{\theta}$ obtained in the first step to compute the new weights in equation (4.10). We resample one observation according to these weights.

For the Wean Hall data, we run this algorithm with two different IS sample sizes $n$. Figures 7 and 8 show two maps obtained for $n = 10$ and $n = 100$, respectively. For each IS sample size, additional observations obtained closely resemble the ones shown here.

For the data collected at the Pontificia Universidad Catolica de Chile, Figure 9 shows the map obtained from raw data, confirming the error accumulation present in odometry. Figure 10 shows the result of the application or our main algorithm to this data.

Figures 7, 8 and 10 show that this algorithm allows the robot to recover from odometry error, aligning the data into the right number of corridors. Comparing these results with the results obtained by our first algorithm, we find that incorporating laser data when sampling locations significantly improves the behavior of the sampling technique. Figures 7 and 8 closely resemble the true Wean Hall map, shown in Figure 3. Figure 10 also resembles the true area navigated by the robot (not shown here).

Our results also show improvement when we draw more observations in the IS step for sampling locations, although even a small sample size, for example, $n = 10$, behaves reasonably well.

The computational complexity of the algorithm is $\mathcal{O}(n \times T)$. The complexity is linear in the size $n$ of the IS step, as each sampled value is processed independently, which allows to an easy parallelization of the algorithm. In particular, we also reduce processing time by using pre-computed weights, when repeated observations are sampled at the re-sampling step of IS.

The complexity is linear in the number of observations in the sample, $T$, as at each point in time, we only update a fixed number of cells in the neighborhood of the sampled location in the map.

We further reduce the computational complexity in our implementation by considering that a robot usually obtains data at a high rate (about 10 measurements per second). We discard redundant information when the robot is static or moving too slowly.

With these simplifications, our algorithm requires about 1 second to compute the weights of the $n$ observations sampled at each point in time, when running in a Pentium processor at 2 GHz.

**6. Contributions and open statistical modeling problems.** In this paper we have presented a fully Bayesian approach to modeling the simultaneous localization and mapping (SLAM) problem for mobile robots in indoor environments with a simplified sensor system. Our methodology utilizes a probabilistic graphical representation for data acquisition and inference that has already been adapted to a variety of other SLAM problems in the robotics literature. We use importance sampling approaches to calculate the posterior distribution and the methodology, while not implementable in anything approximating real time can serve as a baseline for assessing various real-time approximations, such as the particle filtering methods as well as other approaches.

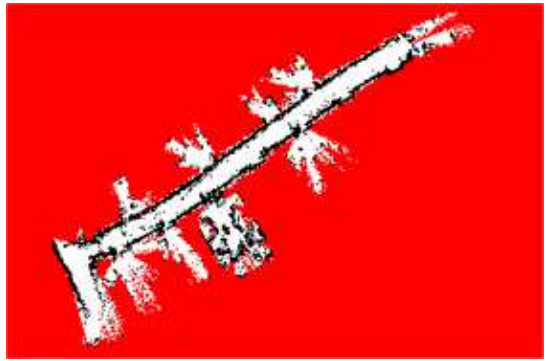

Fig. 7. *Map of Wean Hall, at Carnegie Mellon University, obtained from proposed main algorithm, and IS sample size $n = 10$.*

Finally, our results suggest that modeling odometry error is the key component in modeling the SLAM problem.

Extensions to the models here might well involve distributions over families priors for the maps and over motion and perception models. But to make the methodology truly useful for robotics, we need to explore approximations to the fully Bayesian methods here and compare these to current technologies such as FastSLAM which is built on Kalman filtering ideas and Gaussian approximations. Two approximation approaches seem potentially promising:

1. Variational approximations of the sort that proved extremely valuable in a variety of other machine learning problems, for example, see [14] and [3], that have an EM-like structure.
2. Model approximations emerging from the literature on tropical geometry which have already been applied to simple hidden Markov models [20, 21].

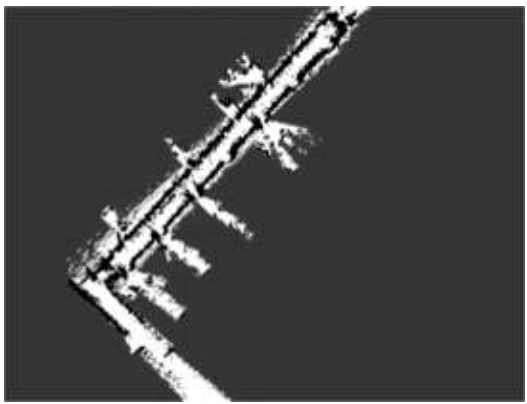

Fig. 8. *Map of Wean Hall, at Carnegie Mellon University, obtained from proposed main algorithm, and IS sample size $n = 100$.*

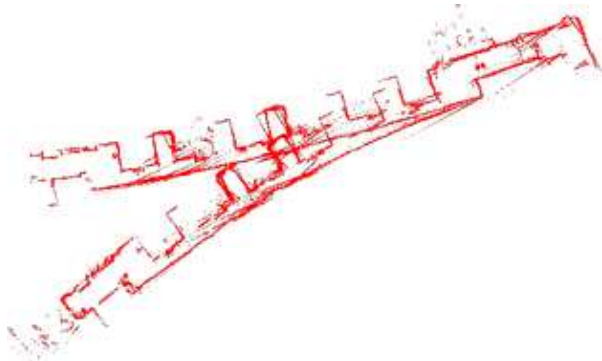

Fig. 9. *Map of the Computer Science Department, at Universidad Catolica de Chile, obtained from raw data. The true area possesses only one corridor.*

Variational approximations should be reasonably easy to implement, while the tropical geometry approaches likely will require new mathematical and statistical work.

As we mentioned at the outset, we have addressed a relatively simple and stylized probabilistic robotics problem. Current sensors in widespread use provide more elaborate information of environments, including pictures, and many of the most interesting environments are three-dimensional [12] and either dynamic or do not have the simple restricted boundaries exhibited by our data and modeling [27]. Other settings involve multiple robots interacting within an environment [9]. These mapping and localization problems provide serious challenges to both machine learning and statistics. For most of these problems, the FastSLAM methodology of Thrun and his colleagues remains the state-of-the-art.

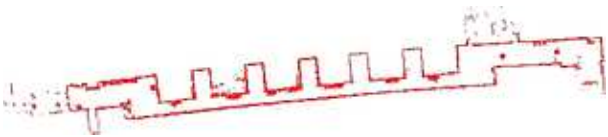

Fig. 10. *Map of the Computer Science Department, at Universidad Catolica de Chile, obtained from proposed main algorithm, and IS sample size $n = 100$.*

## APPENDIX: DETAILS FOR FIRST ALGORITHM

Consider

$$(A.1) \qquad P(\theta_{tk}|\boldsymbol{Z},\boldsymbol{V},\boldsymbol{\theta}^{t-1}) \propto P(\theta_{tk}|V_{tk})\frac{P(\theta_{tk},\boldsymbol{Z},\boldsymbol{\theta}^{t-1},\boldsymbol{V}_{(tk)})}{P(\theta_{tk},\boldsymbol{V}_{(tk)})}.$$

Based on the fact that $\theta_{tk}$ is independent of laser readings that do not contain a reading $k$ at time $t$, $V_{(tk)}$, when no information about locations is available, we can decompose the denominator in equation (A.1) in a product of marginals. We also decompose the term $P(\theta_{tk},\boldsymbol{Z},\boldsymbol{\theta}^{t-1},\boldsymbol{V}_{(tk)})$ in the numerator and, under a Gaussian perception model, we obtain

$$
\begin{aligned}
P(\theta_{tk}|\boldsymbol{Z},\boldsymbol{V},\boldsymbol{\theta}^{t-1}) &\propto P(\theta_{tk}|V_{tk})\frac{P(\boldsymbol{Z},\boldsymbol{\theta}^{t-1},\boldsymbol{V}_{(tk)})P(\theta_{tk}|\boldsymbol{Z},\boldsymbol{\theta}^{t-1},\boldsymbol{V}_{(tk)})}{P(\theta_{tk})P(\boldsymbol{V}_{(tk)})} \\
&= \frac{\phi((V_{tk}-\theta_{tk})/\sigma)}{\Phi((d_{\max}-\theta_{tk})/\sigma)-\Phi(-\theta_{tk}/\sigma)}P(\theta_{tk}|\boldsymbol{Z},\boldsymbol{\theta}^{t-1},\boldsymbol{V}_{(tk)}) \\
&\quad \times I(0<V_{tk}<d_{\max})I(\theta_{tk}>0),
\end{aligned}
$$

where the functions $\phi$ and $\Phi$ are the density and cumulative function of the standard Gaussian distribution, respectively.

**Truncated geometric distribution approximation.** Consider that in the term $P(\theta_{tk}|\boldsymbol{Z},\boldsymbol{\theta}^{t-1},\boldsymbol{V}_{(tk)})$, $\boldsymbol{Z}^{t-1}$ and $\boldsymbol{\theta}^{t-1}$ convey deterministic information about the map. Thus, the sample space of $\theta_{tk}$ gets narrower once $Z_t$ is also known. Let $\boldsymbol{Z}_{t+1}^T$ be the set of locations between times $t+1$ and $T$. Calculation of the term $P(\theta_{tk}|\boldsymbol{Z},\boldsymbol{\theta}^{t-1},\boldsymbol{V}_{(tk)})$ is difficult because $\boldsymbol{Z}_{t+1}^T$ and $\boldsymbol{V}_{(tk)}$ also convey information about the map but, since the laser readings carry error, this information is probabilistic instead of deterministic. Thus, introducing this information involves introducing additional integration. To avoid this, we use

$$P(\theta_{tk}|\boldsymbol{Z},\boldsymbol{\theta}^{t-1},\boldsymbol{V}_{(tk)}) \approx P(\theta_{tk}|\boldsymbol{Z}^{t-1},\boldsymbol{\theta}^{t-1},Z_t).$$

The distribution in this equation takes the value zero for those $\theta_{tk}$ that are inconsistent with $\boldsymbol{Z}^{t-1}$, $\boldsymbol{\theta}^{t-1}$ and $Z_t$.

Figure 11 shows this situation. The dashed cell corresponds to the location of the robot at time $t$, $Z_t$. Using the information conveyed by $\boldsymbol{Z}^{t-1}$ and $\boldsymbol{\theta}^{t-1}$, we label each cell l as either "1" or "0" depending on whether we determine it to be occupied or empty. The arrow corresponds to the direction of the $k$th beam. We see that the closest obstacle in that direction can only be located at the unlabeled cells or at the occupied cell, in the direction of the arrow. We denote the set of these possible cells as $C_{tk}$.

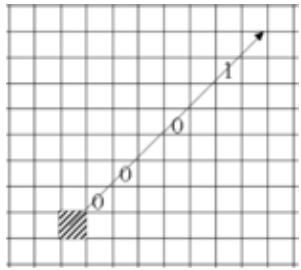

Fig. 11. *Illustration of a truncated geometric distribution.*

On the other hand, the probability that the closest obstacle is in one of the cells in $C_{tk}$ corresponds to a certain particularization of a *Geometric* distribution with parameter $p$, where $p$ is the prior probability of each cell of being occupied. These considerations imply that the distribution $P(\theta_{tk}|\boldsymbol{Z}^{t-1}, \boldsymbol{\theta}^{t-1}, Z_t)$ corresponds to the so-named *Truncated Geometric* distribution, $Tr.Geom(C_{tk}, p)$.

**Acknowledgments.** We thank Sebastian Thrun who introduced us to the SLAM problem, provided feedback on the approach presented here at various stages, and arranged to share with us the Carnegie Mellon data analyzed in this paper. Jay Kadane provided crucial input on the development and implementation of the algorithms. This work is partially based on material from the Ph.D. thesis of the first author in the Department of Statistics at Carnegie Mellon University.

A. ARANEDA
DEPARTMENT OF STATISTICS
PONTIFICIA UNIVERSIDAD CATÓLICA DE CHILE
SANTIAGO
CHILE
E-MAIL: aaraneda@mat.puc.cl

S. E. FIENBERG
DEPARTMENT OF STATISTICS
CARNEGIE MELLON UNIVERSITY
PITTSBURGH, PENNSYLVANIA 15213
USA
E-MAIL: fienberg@stat.cmu.edu

A. SOTO
DEPARTMENT OF COMPUTER SCIENCE
PONTIFICIA UNIVERSIDAD
   CATÓLICA DE CHILE
SANTIAGO
CHILE
E-MAIL: asoto@ing.puc.cl